\documentclass[sigplan, 10pt,nonacm]{acmart}

\usepackage{xurl}
\usepackage{subcaption}
\usepackage{enumitem}
\usepackage{graphicx}
\usepackage{tabularx}
\usepackage{caption}
\usepackage{multirow}
\usepackage{array}
\usepackage{pifont}
\usepackage{algorithm}
\usepackage{algpseudocode}

\usepackage{booktabs,makecell,array}
\newcolumntype{L}[1]{>{\raggedright\arraybackslash}p{#1}}
\newcolumntype{C}[1]{>{\centering\arraybackslash}p{#1}}
\newcolumntype{T}{@{\,\,\,\,}c@{\,\,\(\mid\)\,\,}c@{\,\,\(\mid\)\,\,}c@{\,\,\,\,}}

\newcommand{\myline}[1]{\noindent\textbf{#1.} }

\usepackage[normalem]{ulem}

\usepackage[most]{tcolorbox}

\newtcolorbox{grayblock}{
  colback=gray!15,
  colframe=gray!15,
  boxrule=0pt,
  arc=2pt,
  left=6pt,
  right=6pt,
  top=6pt,
  bottom=6pt
}

\setlist{
  listparindent=\parindent,
  parsep=0pt,
}

\makeatletter
\@ACM@balancefalse
\makeatother

\begin{document}

\title{RiftANN: Efficient Graph Traversal for Vector Search with RDMA-Based Memory Disaggregation}

\author{Qi Lin}
\affiliation{
  \institution{Arizona State University}
  \city{Tempe}
  \state{Arizona}
  \country{USA}
}
\email{qlin36@asu.edu}

\author{Zhenyu Zhang}
\affiliation{
  \institution{Arizona State University}
  \city{Tempe}
  \state{Arizona}
  \country{USA}
}
\email{zzhan641@asu.edu}

\author{Jun Kong}
\affiliation{
  \institution{Arizona State University}
  \city{Tempe}
  \state{Arizona}
  \country{USA}
}
\email{jkong123@asu.edu}

\author{Zhichao Cao}
\affiliation{
  \institution{Arizona State University}
  \city{Tempe}
  \state{Arizona}
  \country{USA}
}
\email{zhichao.cao@asu.edu}

\renewcommand{\shortauthors}{Qi et al.}

\begin{abstract}

Billion-scale vector collections increasingly produce graph indexes that exceed the local DRAM capacity of a single server. Passive RDMA-based disaggregated memory offers scalable capacity without requiring computation at memory nodes, but executing conventional best-first graph search over this architecture introduces substantial overhead. Fine-grained and frequently unnecessary remote reads increase communication cost, while dependencies between candidate evaluation and subsequent expansion serialize traversal and leave CPU resources idle.

We present RiftANN, a graph-based approximate nearest neighbor search system designed for passive RDMA memory pools. RiftANN first uses compact vectors at compute nodes to navigate the graph and selectively verifies exact vectors according to a threshold calibrated from observed approximation errors. A bounded one-sided RDMA pipeline batches and overlaps the remaining remote accesses with local processing. RiftANN further evaluates returned neighbors in parallel and incrementally incorporates completed results. An impact gate estimates whether unfinished evaluations are likely to change the candidate set, allowing traversal to continue when their expected influence is small while preventing unsafe advancement from substantially incomplete search state. Finally, a lightweight feedback controller coordinates RDMA concurrency with background evaluation capacity as search configurations and query loads change.

We evaluate RiftANN on SIFT, DEEP, and SPACEV at 100M scale and on the billion-vector SIFT1B dataset. At matched recall, RiftANN achieves \(1.6\times\)--\(4.3\times\) latency speedups and \(1.1\times\)--\(2.2\times\) higher throughput than DistVS, while providing \(1.6\times\)--\(5.6\times\) latency speedups over the SSD-based DiskANN and PipeANN systems. These results demonstrate that passive disaggregated memory can support low-latency graph-based vector retrieval without relying on query-time computation at memory nodes.

\end{abstract}

\settopmatter{printfolios=true}
\maketitle
\pagestyle{plain}

\section{Introduction}
\label{sec:introduction}

Approximate nearest neighbor search (ANNS) is a fundamental building block in modern AI and data systems, powering applications such as semantic search~\cite{reimers2019sbert}, recommendation~\cite{covington2016youtube}, and retrieval-augmented generation~\cite{lewis2020rag}. 
Among ANNS methods, graph-based approaches are widely used because they provide high recall at low latency~\cite{wang2021graphsurvey,hnsw,vsag}, typically organizing vectors as a graph traversed using best-first search~\cite{pipeann,jayaram2019diskann}.
However, the high performance of graph-based ANNS relies on a systems assumption: the index fits in fast local memory~\cite{hnsw,jayaram2019diskann}. 
However, as vector datasets grow from millions to billions of high-dimensional embeddings, graph indexes can easily reach hundreds of gigabytes or even terabytes~\cite{jayaram2019diskann,pipeann}, exceeding the DRAM capacity of a single server and making high-performance search increasingly difficult to sustain on one machine.

Disaggregated memory (DM) offers a path beyond this capacity limit by decoupling compute (e.g., graph traversal and distance evaluation) from memory (e.g., index storage) over widely deployed high-speed interconnects (e.g., RDMA~\cite{gu2017efficient,aifm}). A natural approach is to retain graph indexes and best-first search procedures (e.g., Vamana~\cite{jayaram2019diskann}) while placing the index in remote memory and executing queries at a compute node. However, our measurements in~\S\ref{sec:simple_solution} show that this straightforward approach increases query latency by \(1.9\times\)--\(2.1\times\) over local execution at matched recall. Remote access overhead explains only part of the gap: the non-RDMA portion of latency also increases from 291 to 374\,\textmu s. 

This slowdown reflects the distinct cost model of RDMA-based DM between local DRAM and SSDs~\cite{pipeann,kalia2016rdma,shan2018legoos}. Local DRAM provides inexpensive random access, making graph search primarily compute-bound~\cite{vsag,hnsw, douze2024faiss, oguri2024entry,li2020adaptiveet}, whereas SSD systems amortize long block I/O through batching, packing, and pipelining~\cite{jayaram2019diskann,starling,pipeann}. RDMA provides byte-addressable access, but fine-grained reads still incur microsecond-scale latency and per-request NIC overhead. On our testbed, a 128~B RDMA read takes 2--3\,\textmu s in isolation and 4--5\,\textmu s under contention. Moreover, best-first search cannot select subsequent expansions until returned candidates are evaluated and incorporated into the search frontier, introducing approximately 69\,\textmu s of CPU idle time per query. Naive remote execution therefore exposes two bottlenecks: high remote-access costs from numerous fine-grained reads and CPU stalls caused by dependent graph search stages.

Existing DM-based ANNS systems address only parts of these costs and follow two DM deployment models. Active-DM equips memory nodes or near-memory endpoints with compute capability, whereas passive-DM exposes stored data through remote reads and performs all query processing at compute nodes. Active-DM systems such as DistVS~\cite{distvs} and CXL-ANNS~\cite{jang2023cxlanns} reduce data transfer through query-time computation near memory, but their serving capacity is constrained by the compute resources provisioned at the memory tier. The passive-DM system like d-HNSW~\cite{dhnsw} keeps query processing at compute nodes and amortizes RDMA overhead through partition-level access, but preserving candidate coverage and recall requires fetching more partitions, weakening the traffic savings. Existing systems therefore either rely on memory-side computation or trade remote traffic for search quality. \textit{\textbf{These limitations motivate us to build a high-performance graph-based ANNS system for passive-DM to reduce both remote-access costs and stalls caused by search dependencies without sacrificing search quality.
}}

Achieving this goal presents three challenges. \underline{\textit{First}}, on the communication side, fine-grained graph accesses incur substantial per-request overhead and collectively generate large volumes of remote traffic, posing the challenge of building an efficient communication substrate over passive-DM that reduces both request overhead and remote traffic while preserving search quality. \underline{\textit{Second}}, on the computation side, the strict dependency between candidate evaluation and subsequent expansion serializes graph traversal and places local computation on the critical path. The challenge is to relax this dependency while preventing incomplete candidate information from misguiding traversal. \underline{\textit{Finally}}, the performance bottleneck shifts between communication and computation as workloads and execution conditions change, requiring adaptive coordination to sustain high throughput and low tail latency without workload-specific tuning.

To address these challenges, we build RiftANN, a high-performance graph-based ANNS system for passive-DM. RiftANN first introduces a \textit{\textbf{Calibrated One-Sided Search Pipeline}} that uses compact local representations to guide graph traversal, avoiding exact-vector reads for most discovered candidates. It then applies a verification rule calibrated from measured approximation errors to fetch exact vectors only for unexpanded candidates that may affect the final top-\(k\) results. The resulting requests retrieve only the required record fields through a unified RDMA path that batches them for asynchronous execution. This design reduces request overhead, remote traffic, and exposed latency while preserving high recall.

RiftANN addresses the compute-side bottleneck through \textbf{\textit{Impact-Aware Asynchronous Traversal}}, which uses impact-aware traversal control to safely overlap parallel neighbor evaluation with graph expansion. A shared background worker pool evaluates returned neighbors, while the main thread incrementally merges completed results and continues traversal whenever the partially updated candidate set remains sufficiently reliable. To assess this reliability, RiftANN combines the recent frequency of candidate-set updates with the number of unfinished evaluations to estimate how many potentially influential updates may remain hidden. This estimate determines whether the main thread proceeds with subsequent expansion or waits for more evaluations to finish, thereby avoiding an unconditional wait-all barrier and reducing compute-side serialization without allowing incomplete candidate information to misguide traversal.

Finally, RiftANN introduces \textbf{\textit{Feedback-Driven Pipeline Balancing}} to coordinate remote fetching and local neighbor evaluation as a producer-consumer pipeline. It measures pipeline imbalance using a per-query defer ratio that relates unfinished evaluation work to recently generated work, then jointly adjusts worker allocation and RDMA concurrency. This feedback limits backlog growth and resource underutilization as search settings and query loads change, sustaining high throughput and low tail latency without workload-specific tuning.

We evaluate RiftANN on three 100M-scale datasets and SIFT1B~\cite{jegou2011sift1b, simhadri2022bigann, babenko2016deep1b}. At 0.90 Recall@10, RiftANN provides \(1.9\)--\(5.6\times\) latency speedups and \(1.1\)--\(4.2\times\) higher throughput than DistVS~\cite{distvs}, PipeANN~\cite{pipeann}, and DiskANN~\cite{jayaram2019diskann} on the 100M-scale datasets. On SIFT1B, it achieves \(1.62\)--\(2.40\times\) latency speedups and \(1.11\)--\(2.47\times\) higher throughput than the same baselines. Compared with graph-matched in-memory Vamana~\cite{jayaram2019diskann} on the 100M-scale datasets, RiftANN incurs \(1.8\)--\(2.7\times\) the latency while supporting remotely stored indexes within a \(24\,\mathrm{GB}\) compute-side memory budget. These results show that passive-DM can support low-latency graph search without query-time computation at memory nodes. RiftANN is designed and evaluated for passive RDMA-based DM. Its compute-side mechanisms may also apply to lower-latency CXL-based memory, although differences in access latency and endpoint capabilities require separate evaluation.

\vspace{-8pt}
\section{Background and Motivations}
\label{sec:background}

\subsection{Graph-Based ANNS and Best-First Search}

\textbf{Graph-Based ANNS.} Given a query vector, $k$-nearest neighbor search identifies the $k$ closest vectors in a dataset. Exact search is prohibitively expensive in high-dimensional spaces~\cite{indyk1998approximate}, motivating approximate nearest neighbor search (ANNS)~\cite{li2020annanalysis} to retrieve approximate top-$k$ results at substantially lower cost. We measure search accuracy using recall, the fraction of the true $k$ nearest neighbors returned. Among ANNS approaches, graph-based methods provide a strong balance between recall and search efficiency~\cite{wang2021graphsurvey,hnsw}. They represent vectors as graph nodes, connect nearby vectors with edges, and answer queries through graph traversal.

\noindent \textbf{Best-First Search.} Graph-based ANNS commonly uses best-first search to traverse the graph and identify approximate top-$k$ neighbors~\cite{pipeann,jayaram2019diskann}. It maintains a bounded candidate set ordered by distance to the query and repeatedly selects one or more closest unexpanded candidates. Processing a selected candidate consists of two logical stages. Expansion accesses the candidate’s adjacency list and exposes previously unseen neighbors, while evaluation computes their distances, inserts competitive neighbors, and prunes the candidate set to retain only the closest candidates. These updates determine which candidates may be expanded next. The search terminates when no unexpanded candidate can improve the current search bound, after which the closest top-$k$ candidates are returned.

Best-first search exhibits two key characteristics. First, conventional execution completes neighbor evaluation and candidate-set updates before selecting subsequent expansions, creating an evaluation-to-expansion ordering dependency. Second, graph traversal generates many fine-grained, random vector accesses (e.g., neighbor vectors). Although a node’s neighbor IDs are stored contiguously, the corresponding vectors are scattered across the index~\cite{hnsw,nsg,jayaram2019diskann}, so exploring a single node requires fetching multiple scattered vectors, often in fine-grained units (e.g., less than 1~KB).

\subsection{In-Memory and SSD-Based ANNS}

The same best-first search procedure exposes different bottlenecks when the index resides in local DRAM or on SSD, leading to distinct optimization strategies.

With the index in local DRAM, fine-grained random accesses are inexpensive and expose little latency across dependent search iterations, making distance computation and traversal major costs. In-memory systems therefore optimize vector processing through SIMD/AVX~\cite{douze2024faiss,vsag} and reduce memory stalls or traversal work through prefetching~\cite{vsag}, improved entry-point selection~\cite{oguri2024entry,vsag}, candidate pruning~\cite{vsag}, and early termination~\cite{li2020adaptiveet}. Graph indexes such as HNSW, NSG, and Vamana further improve connectivity to reduce visited nodes at a target recall~\cite{hnsw,nsg,jayaram2019diskann}.

SSD-based systems support larger indexes but translate fine-grained graph accesses into high-latency random block reads, shifting search to an I/O-bound regime and causing I/O amplification~\cite{jayaram2019diskann,pipeann}. Existing systems reduce disk accesses through graph organization, memory-resident routing, and filtering~\cite{jayaram2019diskann,spann,starling}, amortize each I/O through batching, node packing, and locality-aware layouts~\cite{jayaram2019diskann,starling}, and hide the remaining latency through asynchronous pipelining~\cite{pipeann}.

\begin{figure}
    \centering
    \includegraphics[height=3.cm, width=0.9\linewidth]{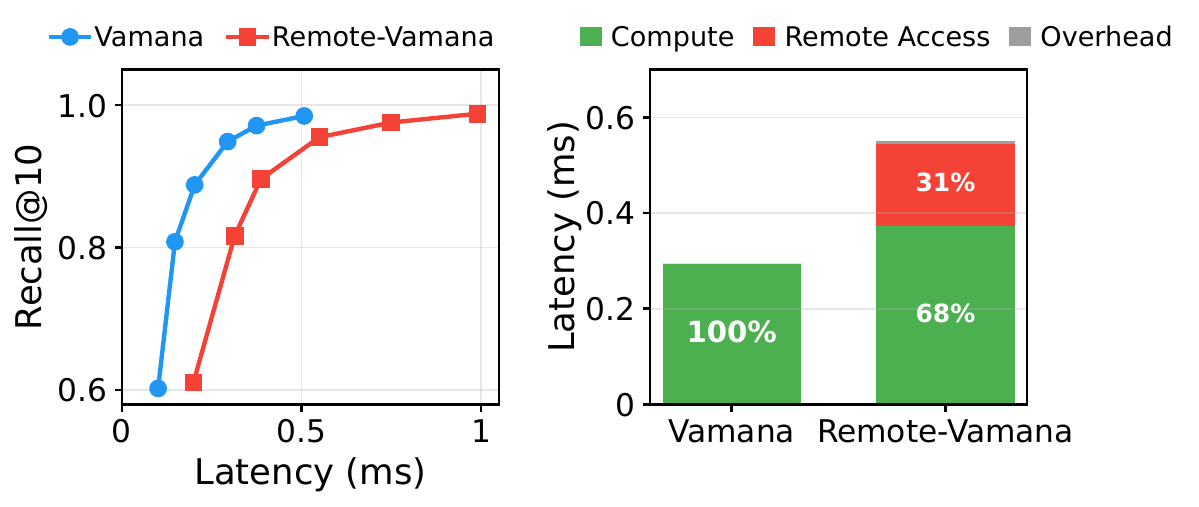}
    \caption{Performance of the naive solution.} 
    \Description{A comparison between an in-memory baseline and a direct RDMA-based extension, highlighting the latency gap.}
    \label{fig:naive_solution}
\end{figure}

\subsection{Motivations}
\label{sec:simple_solution}

Disaggregated memory (DM) decouples compute from memory and exposes a scalable memory pool through RDMA or CXL fabrics~\cite{aifm,jang2023cxlanns,distvs,dhnsw}. It provides capacity beyond local DRAM and finer-grained access than block-oriented SSDs, but retains remote-access latency and per-request overhead, creating a distinct cost model for graph-based ANNS. DM deployments differ in whether the memory tier participates in query execution. Active-DM designs perform operations near memory~\cite{jang2023cxlanns,distvs}, whereas passive-DM designs leave all query processing to compute nodes~\cite{dhnsw}. We target passive RDMA-based DM, where the full graph index and vector dataset reside in remote memory. Compute nodes retain only budgeted auxiliary representations and per-query state while performing all query processing.

A natural and straightforward baseline under this model is to retain an in-memory graph index and its best-first search procedure while replacing local index accesses with one-sided RDMA reads. We instantiate this baseline using Vamana~\cite{jayaram2019diskann} and refer to it as \textit{Remote-Vamana}. To give this baseline a favorable configuration, Remote-Vamana stores high-precision compressed vectors at the CN to guide traversal and minimize remote exact-vector accesses. As shown in~\autoref{fig:naive_solution}, Remote-Vamana incurs \(1.9\times\)--\(2.1\times\) the query latency of in-memory Vamana at matched recall. Its approximately \(170\,\mu\text{s}\) of remote-access time explains only part of this gap, while the non-RDMA portion also increases from \(291\) to \(374\,\mu\text{s}\). These measurements expose two distinct costs. Fine-grained random reads incur fixed RDMA request and completion overhead, while lock-step best-first execution requires the query thread to evaluate returned neighbors and update the candidate set before selecting subsequent expansions, keeping evaluation serialized on the critical path rather than overlapping it with later remote accesses.

\begin{table}[t]
    \centering
    \caption{Sensitivity of existing DM-based ANNS designs.
    C/M denotes CN/MN threads.}
    \label{tab:dm_sensitivity}

    \small
    \renewcommand{\arraystretch}{0.88}
    \setlength{\tabcolsep}{1pt}
    \setlength{\aboverulesep}{1pt}
    \setlength{\belowrulesep}{1pt}

    \begin{tabular}[t]{@{}lcc@{}}
        \toprule
        \multicolumn{3}{c}{\textit{DistVS (R@10 = 0.90)}} \\
        Setting & C/M & KQPS \\
        \midrule
        Original & 26/6  & 13.8 \\
        More CN  & 32/6  & 15.2 \\
        More MN  & 26/12 & 22.4 \\
        \bottomrule
    \end{tabular}%
    \hspace{2pt}%
    \begin{tabular}[t]{@{}lccc@{}}
        \toprule
        \multicolumn{4}{c}{\textit{d-HNSW: routed partitions}} \\
        Setting & \(P/P_{0.90}\) & R@10 & Lat. (ms) \\
        \midrule
        Fewer partitions & 0.50 & 0.84 & 350 \\
        Target recall& 1.00 & 0.90 & 620 \\
        More partitions & 1.25 & 0.91 & 815 \\
        \bottomrule
    \end{tabular}
\end{table}

However, existing systems address only portions of the aforementioned problems. Active-DM systems reduce data transfer through memory-side computation: DistVS~\cite{distvs} places high-precision distance evaluation and search state at memory servers, while CXL-ANNS~\cite{jang2023cxlanns} evaluates distances at compute-capable CXL endpoints. However, this design places memory-side computation on the query-serving critical path. As shown in~\autoref{tab:dm_sensitivity}, on SIFT100M at 0.90 Recall@10, DistVS with 26 CN threads and 6 MN threads saturates at 13.8~KQPS, with 98\% MN-thread utilization. Adding 6 CN threads increases throughput only to 15.2~KQPS, whereas adding 6 MN threads raises it to 22.4~KQPS. This asymmetric improvement identifies memory-side processing as the bottleneck and couples serving capacity to MN compute resources.

The passive-DM system, d-HNSW~\cite{dhnsw}, amortizes RDMA overhead by transferring selected sub-HNSWs as partitions. Here, \(P_{0.90}\) denotes the number of partitions routed per
query to reach 0.90 Recall@10. As shown in~\autoref{tab:dm_sensitivity}, under our 24~GB compute-side memory budget, increasing the routed partitions raises Recall@10 from 0.84 to 0.91 but more than doubles mean latency from \(350\) to \(815\,\mathrm{ms}\), while reaching 0.90 recall already requires approximately \(620\,\mathrm{ms}\). At this operating point, partition transfer, deserialization, and sub-HNSW search account for 82\% of the latency. Partition-granular access therefore amortizes RDMA requests at the cost of transferring and searching more remote data, coupling recall with latency. 

In general, \textbf{\textit{these results motivate us to redesign best-first graph search for passive-DM to require no query-time computation at memory nodes, reduce remote-access overhead and serialization stalls, and preserve high search quality.}}

\subsection{Challenges}
\label{sec:challenges}
Achieving this goal presents three key challenges.

\begin{figure}
    \centering
    \includegraphics[width=0.9\linewidth]{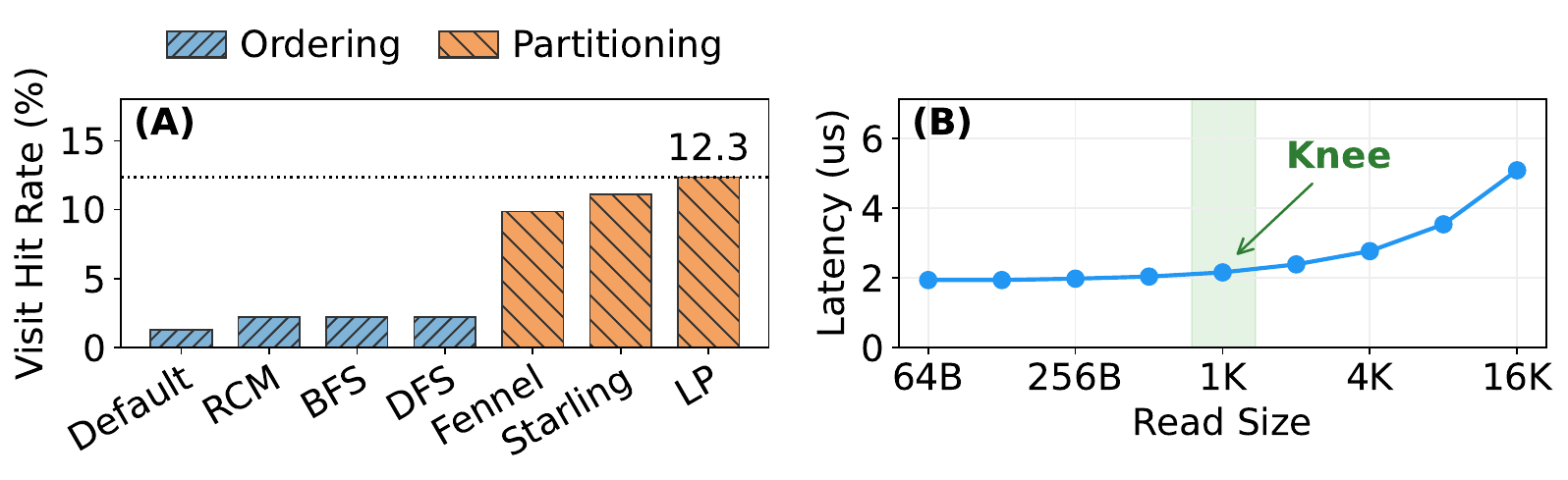}
    \caption{Limits of coarser-grained fetching over RDMA. (a)~Visit hit rate of co-packed nodes. (b)~RDMA read latency versus read size.}
    \Description{Left: bar chart showing that ordering layouts reach 1--2\% visit hit rate and partitioning layouts reach at most 12.3\%. Right: line plot showing RDMA read latency near 2 microseconds up to about 1 kilobyte, rising to about 5 microseconds at 16 kilobytes.}
    \label{fig:packing_waste}
\end{figure}

\noindent \textbf{Challenge 1. Reducing Remote-Access Costs under Limited Local Memory.}
Best-first search exposes neighboring node IDs whose vectors are scattered across the remote index, generating many fine-grained RDMA reads of \(128\)--\(512\,\mathrm{B}\), depending on vector dimensionality and data type~\cite{dhnsw}. Coarser-grained fetching can amortize per-request overhead, but partition-level access, as in d-HNSW, restricts candidate coverage. A less restrictive alternative is to co-locate nodes likely to be visited together during graph-wide traversal. However,~\autoref{fig:packing_waste}(a) shows that representative ordering and partitioning schemes (e.g., Fennel~\cite{tsourakakis2014fennel}, Starling~\cite{starling}, and LP~\cite{ugander2013blp}) achieve a visit hit rate of at most 12.3\%, measured as the fraction of co-packed nodes subsequently visited by the same query. Meanwhile,~\autoref{fig:packing_waste}(b) shows that read latency increases rapidly beyond approximately \(1\,\mathrm{KB}\). Coarser-grained fetching therefore amortizes request overhead only by transferring mostly unused data at increasing latency.

Fine-grained reads avoid spatial overfetching but retrieve the exact vector of every newly discovered neighbor before determining whether it should enter the candidate set. On SIFT100M~\cite{jegou2011sift1b}, increasing the graph degree from 32 to 128 raises exact-vector reads from 2,134 to 4,566 per query, yet 85.4\%--90.4\% of the fetched candidates are never expanded. These reads cannot simply be eliminated because their distances may still affect candidate ranking, pruning, and final top-\(k\) results. Compact representations such as product quantization (PQ)~\cite{jegou2011pq} and scalar quantization (SQ)~\cite{douze2024faiss} enable local distance estimates without fetching every exact vector~\cite{vsag,douze2024faiss,jayaram2019diskann}, but introduce a memory-accuracy trade-off. More accurate representations require greater per-vector storage, while smaller representations may distort traversal and result selection. Under our passive-DM model, limited compute-side memory imposes a bounded per-vector budget, constraining the precision of representations retained for the full dataset. \textit{The challenge is therefore how to limit RDMA request overhead and transferred data within a bounded compute-side memory budget while preserving high recall.}

\begin{figure}
    \centering
    \includegraphics[width=0.9\linewidth]{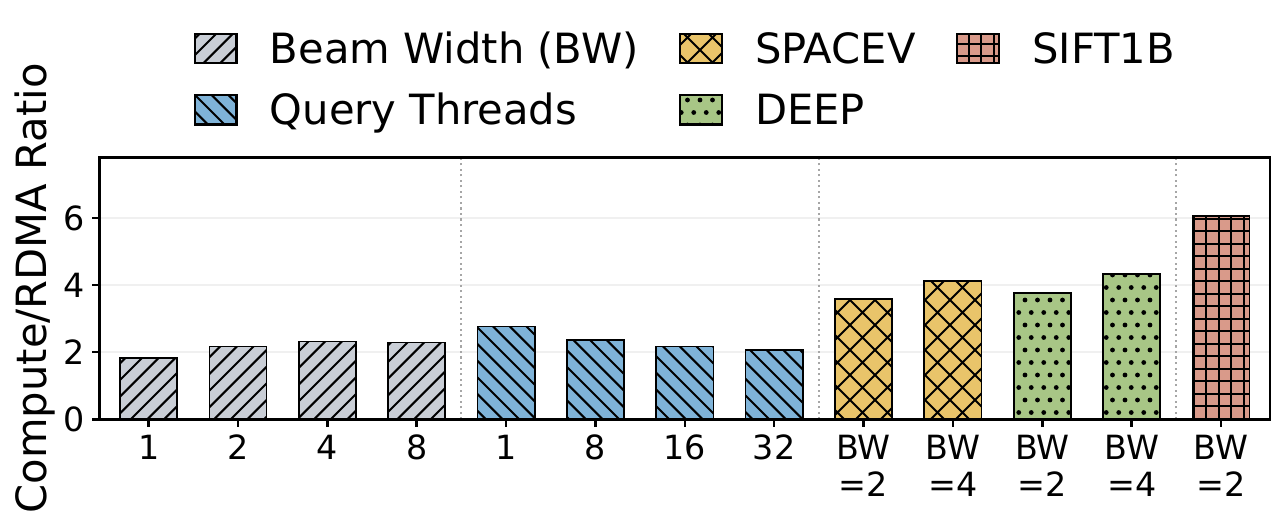}
    \caption{Pipeline balance under different settings.} 
    \Description{different factors will affect the pipeline}
    \label{fig:pipeline_imbalanced}
\end{figure}

\myline{Challenge 2. Overlapping Neighbor Evaluation and Graph Expansion}
Even after reducing remote-access costs, handling the remaining accesses synchronously can stall graph search. We implement an asynchronous Remote-Vamana, termed \textit{Async-Vamana}, which retains the same high-precision local compressed vectors and remote-access decisions but processes completed records while other one-sided RDMA reads remain in flight. On SIFT100M at matched recall, Async-Vamana reduces mean query latency from approximately \(550\,\mu\text{s}\) to \(368\,\mu\text{s}\), a \(1.5\times\) speedup. However, asynchronous RDMA hides remote-access latency without relaxing the evaluation-to-expansion dependency. The query thread still incorporates evaluated neighbors into the candidate set before selecting subsequent expansions. Our profiling shows that candidate evaluation consequently becomes the dominant exposed cost, accounting for approximately 78\% of Async-Vamana's execution time.

Parallelizing candidate evaluation can reduce this serialization but creates a synchronization dilemma. Waiting for all evaluations preserves a fully updated candidate set but retains a wait-all barrier, whereas proceeding earlier exposes traversal to incomplete candidate information. To quantify this risk, we delay each completed evaluation result by one expansion hop before inserting it into the candidate frontier, while keeping the search budget and all other settings unchanged. On SIFT100M, this delay reduces Recall@10 from 99.47\% to 66.56\%. Because best-first traversal is path-dependent, a temporarily hidden promising candidate can alter both the next expansion and the graph region reached within the search budget. Parallel evaluation therefore creates a tension between removing the wait-all barrier and maintaining a sufficiently complete search frontier.

\begin{figure*}
    \centering
    \includegraphics[width=1\textwidth]{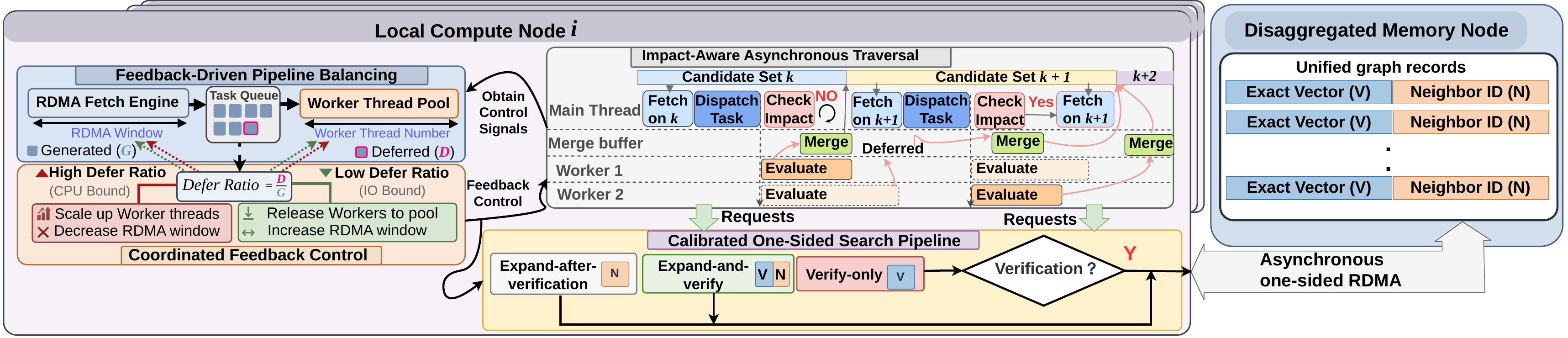}
    \caption{RiftANN Overview.} 
    \Description{}
    \label{fig:overview}
\end{figure*}

Prior systems relax related dependencies at different execution boundaries. CoTra~\cite{zhi2025towards} tolerates delayed candidate information through periodic coordination during distributed traversal, but it assumes compute-active servers and does not address the evaluation-to-expansion dependency in passive-DM. PipeANN~\cite{pipeann} bounds missing neighbor information using pipeline width, but the amount of unfinished work does not indicate whether it may change the search frontier. CXL-ANNS~\cite{jang2023cxlanns} overlaps candidate maintenance with near-memory distance computation, while graph expansion still depends on the unfinished evaluations that determine node selection. \textit{The challenge is therefore how to overlap neighbor evaluation and graph expansion, reducing compute-side serialization while preserving high recall.}

\myline{Challenge 3. Maintaining Pipeline Balance under Dynamic Load}
Once candidate evaluation and remote access are highly parallelized, pipeline efficiency depends on their relative pressure. As shown in~\autoref{fig:pipeline_imbalanced}, the ratio of candidate-evaluation time to RDMA fetching time ranges from \(1.8\times\) to \(6.1\times\) across datasets, search configurations (e.g., different beam widths), and query-concurrency levels. A configuration tuned for one setting can therefore become unbalanced in another, causing either evaluation backlogs or idle compute resources and reducing effective overlap. The balance can shift even under the same configuration. As detailed in \S\ref{sec:feedback_analysis}, temporarily doubling the query arrival rate under a fixed configuration causes the evaluation backlog to grow, increasing P99 latency from \(1.46\,\text{ms}\) to \(3.02\,\text{ms}\). \textit{The challenge is therefore how to maintain pipeline balance as computation and communication pressures shift, sustaining high throughput and low tail latency without workload-specific tuning}.

\vspace{-5pt}
\section{The Design of RiftANN}
\label{sec:design}

\subsection{Overview}

We build RiftANN, an end-to-end graph-based ANNS system for passive disaggregated memory. As shown in~\autoref{fig:overview}, passive memory nodes store unified graph records containing each node's exact vector and adjacency list. On compute nodes, a per-query main thread drives a closed-loop search. Traversal generates state-dependent expansion and verification reads through the calibrated pipeline. Returned exact vectors update the exact-result set, while returned adjacency lists generate evaluation tasks for the shared worker pool. The main thread incrementally merges completed evaluations, and the impact gate determines whether subsequent expansion may proceed. A feedback controller jointly adjusts the RDMA window and worker allowance as pipeline pressure changes. RiftANN comprises three components:

\myline{Calibrated One-Sided Search Pipeline}
Compact local representations guide traversal. Candidate state determines the required record fields, while an error-calibrated rule selects unexpanded candidates for exact verification. A bounded asynchronous RDMA path batches and issues the resulting reads, reducing request overhead, transferred data, and exposed remote latency.

\myline{Impact-Aware Asynchronous Traversal}
Shared workers evaluate newly exposed neighbors while the main thread incrementally merges completed results. The impact gate estimates the candidate-set updates hidden by unfinished work and admits subsequent expansion only when this estimate remains within a safety budget, avoiding unconditional wait-all stalls while limiting candidate-set incompleteness.

\myline{Feedback-Driven Pipeline Balancing}
A per-query defer ratio relates unfinished neighbor evaluation work to recent task generation. The controller uses this signal to jointly adjust worker allowance and RDMA concurrency, maintaining effective overlap as search demand and query load change.

\subsection{Calibrated One-Sided Search Pipeline}
\label{sec:calibrated}

\begin{figure}
    \centering
    \includegraphics[width=1\linewidth]{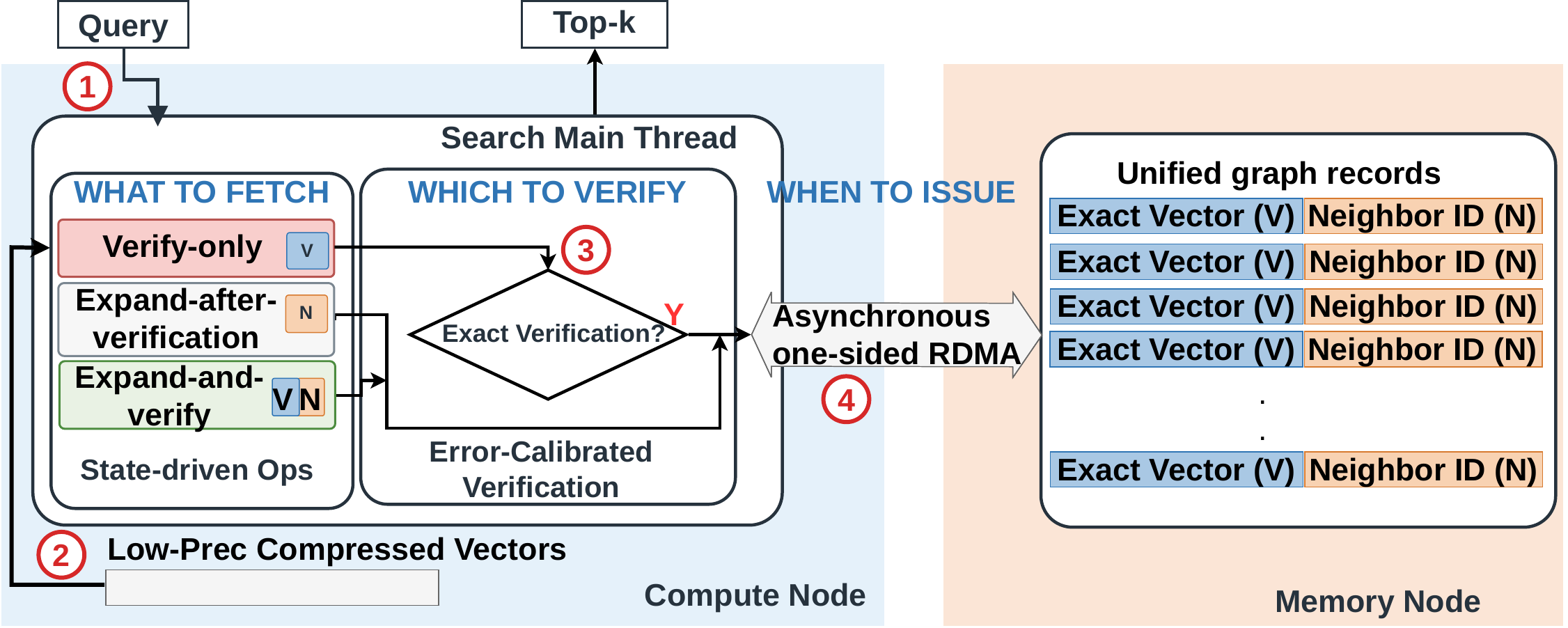}
    \caption{Calibrated One-Sided Search Pipeline.} 
    \Description{}
    \label{fig:design1}
\end{figure}

As shown in~\autoref{fig:design1}, RiftANN coordinates three remote-access decisions: what fields to fetch from each remote record, which unexpanded candidates require exact verification, and when to issue the resulting reads. These decisions are driven by candidate state, calibrated approximation errors, and a bounded asynchronous RDMA pipeline, respectively.

\myline{State-Driven Remote Operations}
Each remote graph record stores an exact-vector prefix followed by an adjacency-list suffix. Expansion requires the adjacency list to continue traversal, while verification requires the exact vector for final top-\(k\) selection. Based on the required operation and previously retrieved data, RiftANN supports three read modes. \textit{Expand-and-verify} fetches both fields when an unverified candidate is selected for expansion. \textit{Verify-only} fetches only the exact vector for an unexpanded candidate admitted by the verification gate. \textit{Expand-after-verification} fetches only the adjacency list when a verified candidate is later expanded. These modes retrieve only the currently required fields and reuse previously fetched data, avoiding speculative cross-node packing and repeated exact-vector transfers.

\myline{Error-Calibrated Exact Verification}
RiftANN uses compact local representations to estimate candidate distances for graph navigation, while exact distances determine the final top-$k$ results. Candidates selected for expansion obtain their exact vectors through \textit{Expand-and-verify}, but discovered candidates that are not expanded may also belong to the final result and require \textit{Verify-only}. Verifying all such candidates would eliminate the traffic savings, while verifying too few could reduce recall. RiftANN therefore uses measured approximation errors to calibrate which unexpanded candidates require exact verification.

Once RiftANN has obtained exact distances for at least \(k\) candidates, let \(\tau\) denote the current \(k\)-th smallest exact distance. Because the verified candidates form only a subset of the dataset, \(\tau\) upper-bounds the ground-truth \(k\)-th distance. Every discovered true top-\(k\) candidate therefore satisfies \(d(v,q)\leq\tau\). Before \(\tau\) is available, RiftANN obtains exact vectors only through \textit{Expand-and-verify}. If approximate and exact distances were identical, verifying unexpanded candidates with \(\hat d(v,q)\leq\tau\) would be sufficient. In practice, approximation errors may place a result-relevant candidate outside this boundary. RiftANN therefore relaxes the boundary by a calibrated factor $\alpha$ and performs \textit{Verify-only} when
\begin{equation}
\hat d(v,q)\leq\alpha\tau.
\end{equation}

RiftANN calibrates \(\alpha\) offline using held-out queries. It runs the same approximate traversal used at runtime and records \(\hat d(v,q)/d(v,q)\) for every discovered candidate that belongs to the ground-truth top-\(k\). Given a target verification miss rate \(\delta\), RiftANN sets
\begin{equation}
\alpha =
\max\left\{
1,\,
Q_{1-\delta}\left(
\frac{\hat{d}(v,q)}{d(v,q)}
\right)
\right\}
\end{equation}
where \(Q_{1-\delta}\) is the empirical \((1-\delta)\)-quantile of the recorded ratios. The outer maximum prevents the verification boundary from becoming narrower than \(\tau\). This calibration is performed once for each dataset and compact-representation configuration.

\textit{Example.} Suppose \(\delta=0.01\) and the 99th percentile of the recorded \(\hat d(v,q)/d(v,q)\) ratios is 1.08. RiftANN sets \(\alpha=1.08\), meaning that at least 99\% of discovered ground-truth top-\(k\) candidates in the held-out queries satisfy \(\hat d(v,q)\leq1.08d(v,q)\). If the current threshold is \(\tau=100\), the runtime verification boundary is 108. A candidate \(v_1\) with \(\hat d(v_1,q)=105\) triggers \textit{Verify-only}, and a later expansion fetches only its adjacency list through \textit{Expand-after-verification}. A candidate \(v_2\) with \(\hat d(v_2,q)=112\) triggers no read unless expanded, in which case \textit{Expand-and-verify} fetches both fields.

For any discovered true top-\(k\) candidate satisfying the calibrated error bound,
\begin{equation}
\hat d(v,q)\leq\alpha d(v,q)\leq\alpha\tau,
\end{equation}
so the candidate passes the runtime verification gate. Selecting the smallest qualifying \(\alpha\) avoids widening the gate beyond the empirical coverage target. As closer candidates are verified, \(\tau\) decreases and the gate becomes progressively more selective. This guarantee is conditional on candidate discovery. Graph traversal determines which candidates are discovered, while calibration determines which discovered candidates receive exact verification. We evaluate verification coverage, recall impact, and exact-vector-read reduction on disjoint test queries in~\S\ref{sec:eval_calibrated_pipeline}.

\myline{Asynchronous One-Sided Execution}
To implement the final decision in~\autoref{fig:design1}, RiftANN issues remote reads through a bounded per-query window that keeps at most \(W_R\) one-sided RDMA reads in flight. Returned exact vectors update the exact-result set and \(\tau\), while returned adjacency lists expose new neighbors for graph traversal. As request slots become available, the current search state determines which expansion and verification reads should be issued.

Because the search state may change while a request is waiting for issuance, RiftANN revalidates each request immediately before posting it. Requests that are no longer needed are discarded, while expansion and verification requests for the same candidate are combined into a single full-record read. The remaining requests are batched using standard RDMA optimizations (e.g., doorbell chaining and selective signaling).

Search terminates only when the navigation condition is satisfied, no unverified candidate remains that satisfies the calibrated gate, and no outstanding read can change the navigation or exact-result state. Although this asynchronous execution hides much of the latency, the main thread still evaluates returned neighbors and incorporates their results before selecting subsequent expansions. Once remote access is largely overlapped, this evaluation-to-expansion dependency becomes the next critical-path bottleneck.

\begin{figure}
    \centering
    \includegraphics[width=\linewidth]{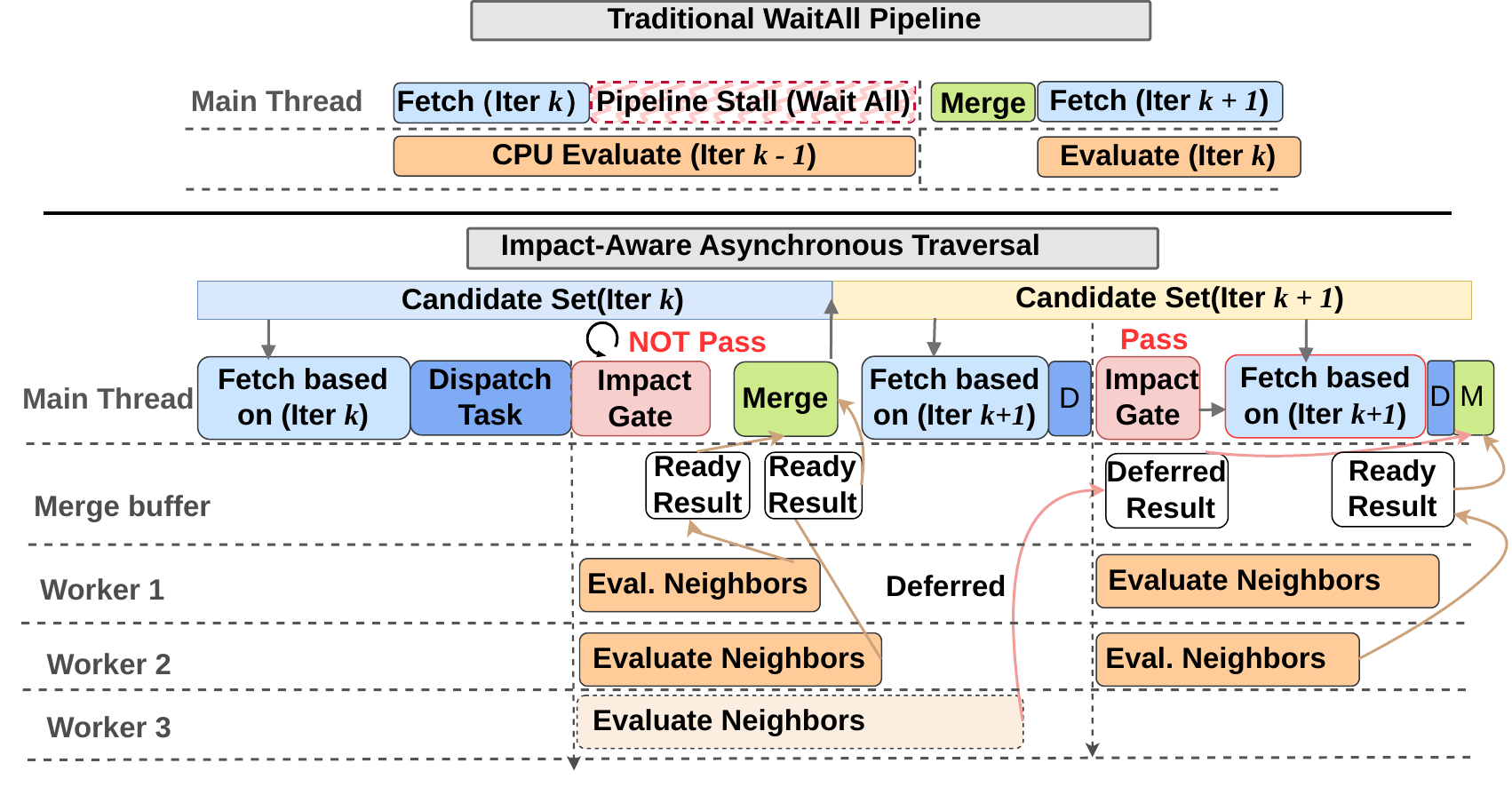}
    \caption{
Pipeline execution under WaitAll and impact-aware traversal. WaitAll blocks subsequent expansion until all evaluations complete. RiftANN incrementally incorporates completed results and admits the next expansion when the impact gate opens, while the remaining evaluation tail continues in parallel.
  } 
    \Description{.}
    \label{fig:lazy_merge}
\end{figure}

\vspace{-5pt}
\subsection{Impact-Aware Asynchronous Traversal}
\label{sec:impact_aware_traversal}

Impact-Aware Asynchronous Traversal overlaps candidate evaluation with subsequent expansion while constraining the estimated impact of unfinished evaluations on the candidate set. A shared background worker pool evaluates returned neighbors, while the main thread incrementally merges completed results. Before each expansion, RiftANN estimates how many candidate-set updates may remain hidden and proceeds only when this estimate is sufficiently small. This state-dependent control removes the unconditional wait-all barrier while preventing unrestricted traversal from an incomplete candidate set.

\myline{Estimating Hidden Candidate-Set Updates}
An unfinished evaluation can affect traversal only if it inserts its candidate into the candidate set, but this likelihood changes substantially during a query. On SIFT100M, the mean candidate-update ratio decreases from 41.0\% over the first 20\% of processed nodes to 0.23\% over the final 40\%, even though the latter stage still evaluates 64.8 neighbors per processed node on average. The unfinished backlog therefore measures the amount of pending work, but not its likely impact on the candidate set. Bounding only the backlog can unnecessarily restrict evaluation--expansion overlap as the search converges. Because the unfinished evaluations and the completed evaluations in the rolling window originate from recent expansions, RiftANN uses the recent candidate-update ratio as an online proxy for their update likelihood at the current search stage. It counts every insertion because the inserted candidate may affect a later expansion decision even if it is not selected next. RiftANN then combines this ratio with the current unfinished-evaluation backlog to estimate how many candidate-set updates may remain hidden from traversal.

Formally, let \(\mathcal{W}_q\) contain the \(H\) most recently completed evaluations for query \(q\), and let \(x_i=1\) if evaluation \(i\) inserts its candidate into the candidate set and \(x_i=0\) otherwise. RiftANN conservatively sets \(\rho_q=1\) until \(H\) evaluation results are available. It then computes the candidate-update ratio as
\begin{equation}
\rho_q=\frac{1}{H}\sum_{i\in\mathcal{W}_q}x_i.
\end{equation}

Let \(U_q\) denote the number of individual candidate evaluations submitted to background workers whose results have not yet been returned to the main thread. After incorporating all currently completed results, RiftANN estimates the number of candidate-set updates still hidden from traversal:
\begin{equation}
\Psi_q=\rho_q U_q.
\end{equation}

Consequently, the same unfinished backlog produces a smaller impact estimate as candidate updates become less frequent, rather than being treated as equally consequential throughout the search.

\myline{Impact-Aware Expansion}
As illustrated in~\autoref{fig:lazy_merge}, a wait-all pipeline blocks subsequent expansion until all evaluations generated by the current expansion have completed. RiftANN instead incorporates evaluation results as they become available and, before each expansion decision, recomputes \(\rho_q\), \(U_q\), and \(\Psi_q\). It admits the next expansion only when:
\begin{equation}
\Psi_q \leq \epsilon
\quad\text{and}\quad
U_q \leq U_{\max}.
\end{equation}
Here, \(\epsilon\) bounds the estimated number of candidate-set updates hidden from traversal, while \(U_{\max}\) provides a hard cap on the unfinished backlog.

\textit{Example.} Consider the default setting \(H=64\), \(\epsilon=1\), and \(U_{\max}=128\). Suppose 16 of the 64 most recently completed evaluations inserted candidates into the candidate set, giving \(\rho_q=0.25\). If \(U_q=8\) evaluations remain unfinished, then \(\Psi_q=0.25\times8=2\). The gate therefore follows the Not Pass branch in~\autoref{fig:lazy_merge}, and the main thread continues merging completed results without starting another expansion. After four of these evaluations complete, suppose the updated ratio remains \(0.25\). The remaining backlog becomes \(U_q=4\), giving \(\Psi_q=1\), so the gate opens. The next expansion can then proceed while the remaining four evaluations continue in parallel. Near convergence, if only 8 of the recent 64 evaluations updated the candidate set, the same backlog of \(U_q=8\) gives \(\Psi_q=1\) and can pass immediately.


Completed results are never intentionally withheld and are incorporated before each expansion decision. If either gate condition fails, the main thread pauses new expansions and continues merging results until the gate reopens. Before terminating, RiftANN drains all unfinished evaluations and reevaluates the navigation and verification conditions, allowing any newly inserted candidate to resume traversal. Unlike the calibrated verification gate, which determines whether an exact vector should be fetched, the impact gate determines only when the next graph expansion may proceed. All completed evaluation results are eventually incorporated.

By regulating expansion with \(\Psi_q\) rather than treating every unfinished evaluation equally, RiftANN bounds the estimated number of hidden candidate-set updates while exposing more evaluation--expansion overlap as the search stabilizes. This reduces the serialized evaluation dependency without unconditionally advancing from an incomplete candidate set. The resulting execution forms a producer--consumer pipeline in which completed expansion reads generate neighbor evaluation tasks and shared workers process them. Because the relative rates of these stages vary with workload and query concurrency, fixed resource allocation can cause evaluation backlogs or worker underutilization, motivating the feedback-driven control described next.

\begin{figure}
    \centering
    \includegraphics[width=1\linewidth]{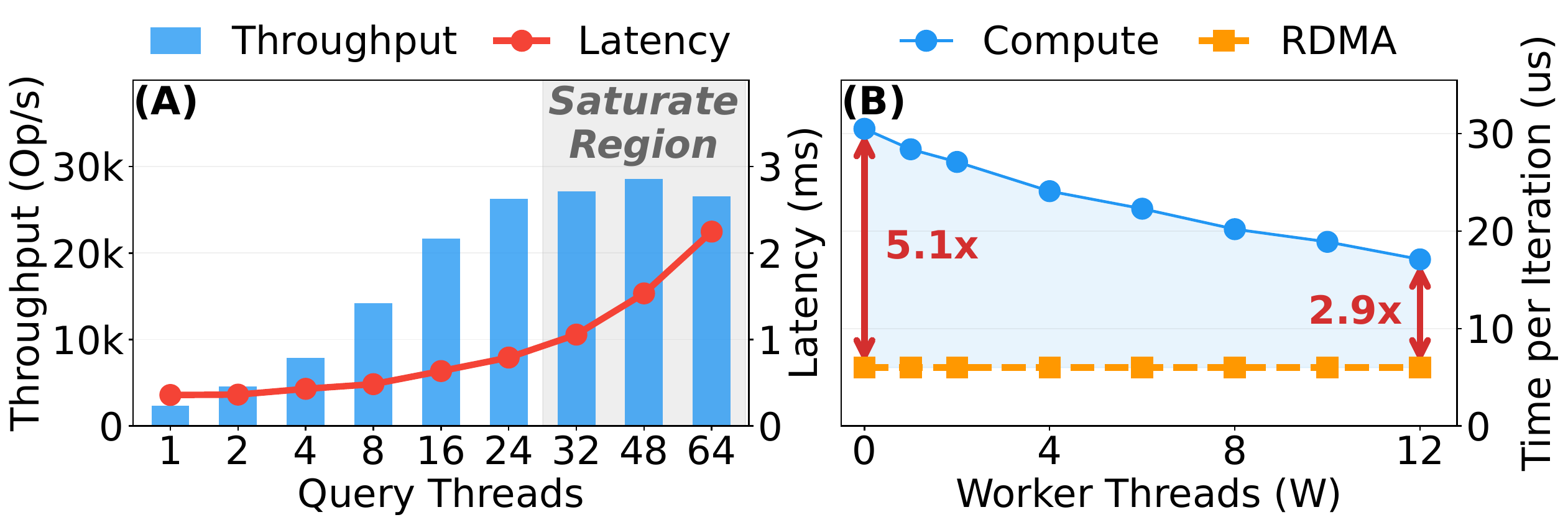}
    \caption{Thread rebalancing between query processing and background workers.} 
    \Description{}
    \label{fig:fig_thread_scaling_balance_workers}
\end{figure}

\subsection{Feedback-Driven Pipeline Balancing}
\label{sec:self-balanced_pipeline}
RiftANN treats remote fetching as the producer and candidate evaluation as the consumer of a per-query pipeline. The RDMA window \(W_{R,q}\) controls task generation, while the worker allowance \(T_q\) controls task processing. A per-query defer ratio coordinates the two to limit backlog growth and worker underutilization as execution conditions change.

\myline{CPU Budgeting across Query and Evaluation Parallelism}
RiftANN operates within a fixed compute-node CPU budget. Each query is driven by one main thread, while candidate evaluations are executed by a background worker pool shared across queries at this CN. As shown in~\autoref{fig:fig_thread_scaling_balance_workers}(a), increasing query concurrency initially improves throughput but eventually yields diminishing returns as CPU, memory, and RDMA resources become contended. Meanwhile, ~\autoref{fig:fig_thread_scaling_balance_workers}(b) shows that assigning compute capacity to background workers substantially reduces candidate-evaluation time while leaving RDMA time largely unchanged. RiftANN therefore bounds the number of active main threads and assigns each query \(q\) a worker allowance \(T_q\), which limits how many shared workers may concurrently process its tasks, subject to pool availability, without reserving physical threads.

\myline{Per-Query Feedback Signal}
Within this fixed CPU organization, RiftANN coordinates two per-query control knobs. The outstanding request window \(W_{R,q}\) bounds the number of one-sided RDMA reads in flight and thereby regulates the rate at which completed expansions generate evaluation tasks. The worker allowance \(T_q\) bounds the number of shared workers that may concurrently process those tasks, regulating their consumption rate. To measure the relative pressure between these two stages, RiftANN computes the defer ratio
\begin{equation}
r_q =
\frac{N_{\mathrm{deferred},q}}
{\max(N_{\mathrm{generated},q},1)}.
\end{equation}

Here, \(N_{\mathrm{deferred},q}\) is the number of evaluation tasks that remain queued or running, while \(N_{\mathrm{generated},q}\) is the number generated since the previous control point. RiftANN invokes the controller after a fixed number of expansion completions and resets \(N_{\mathrm{generated},q}\) after each decision. For example, if 20 tasks have been generated and 5 remain unfinished, then \(r_q=0.25\). Normalizing by recent task generation makes the signal less sensitive to search configuration, RDMA concurrency, and query load. Unlike \(\Psi_q\), which determines whether traversal may advance with unfinished evaluations, \(r_q\) is a performance-control signal used only to adjust worker capacity and outstanding RDMA concurrency.

\begin{algorithm}[t]
\footnotesize
\caption{Feedback-Driven Pipeline Balancing}
\label{alg:self_balance_alg}
\begin{algorithmic}[1]
\State \textbf{Input:} query $q$, $N_{\text{deferred},q}$,
       $N_{\text{generated},q}$, $W_{R,q}$, $T_q$
\State \textbf{Shared:} worker pool $\mathcal{P}$
\State $r_q \leftarrow
       \frac{N_{\text{deferred},q}}
       {\max(N_{\text{generated},q},1)}$
       \Comment{relative computation backlog}

\If{$r_q > \theta_{\mathrm{high}}$}
    \Comment{computation pressure}
    \If{$T_q < T_{\max}$ and shared-pool capacity is available}
        \State $T_q \leftarrow T_q + 1$
        \Comment{grant one worker slot}
    \Else
        \State $W_{R,q} \leftarrow
               \max(W_{R,q}-1, W_{R,\min})$
        \Comment{throttle producer}
    \EndIf

\ElsIf{$r_q < \theta_{\mathrm{low}}$}
    \Comment{low computation pressure}
    \If{$T_q > T_{\min}$}
        \State $T_q \leftarrow T_q - 1$
        \Comment{return capacity to shared pool}
    \Else
        \State $W_{R,q} \leftarrow
               \min(W_{R,q}+1, W_{R,\max})$
        \Comment{increase producer rate}
    \EndIf

\Else
    \State no adjustment
    \Comment{balanced zone}
\EndIf
\end{algorithmic}
\end{algorithm}


\myline{Coordinated Feedback Control}
Rather than targeting a single \(r_q\), RiftANN defines a balanced zone using \(\theta_{\mathrm{low}}\) and \(\theta_{\mathrm{high}}\). It leaves \(W_{R,q}\) and \(T_q\) unchanged within this zone, providing hysteresis against short-lived fluctuations. RiftANN uses the same thresholds and event-based control interval across workloads, with their sensitivity evaluated in~\S\ref{sec:feedback_analysis}.

Algorithm~\ref{alg:self_balance_alg} summarizes the policy. When \(r_q\) exceeds the balanced zone, RiftANN grants additional worker capacity when available and otherwise reduces the RDMA window. When \(r_q\) falls below the zone, it releases workers before increasing RDMA concurrency. This ordering prioritizes compute reallocation before changing work generation. Query bursts naturally raise \(r_q\) and trigger the same feedback loop without a separate detector, while the adjustments reverse as the burst subsides.

\vspace{-5pt}
\section{Evaluation}
\label{sec:evaluation}
We compare RiftANN with state-of-the-art DM- and SSD-based ANNS systems across datasets, quantify the effectiveness, incremental contribution, and trade-offs of its designs, and evaluate scalability across compute and memory nodes.


\subsection{Experimental Setup}
\label{sec:setup}

\noindent\textbf{Basic configuration.}
Unless otherwise stated, experiments use two servers connected by 100-Gbps Mellanox ConnectX-5 RNICs. Each server has two 28-core Intel Xeon Gold 6330 CPUs at 2.00\,GHz and runs Ubuntu 22.04. One server acts as the MN with \(768\,\mathrm{GB}\) of DRAM; the other acts as the CN with \(128\,\mathrm{GB}\) of DRAM and a 2-TB Samsung 980 PRO SSD. The MN stores the full index, while the CN is limited to \(24\,\mathrm{GB}\) for all local data and runtime state. This budget is applied uniformly and is comparable to the \(22.3\)--\(25.0\,\mathrm{GB}\) per-CN footprints reported by DistVS~\cite{distvs}. RiftANN stores a 20-byte PQ code per vector at the CN, requiring approximately \(2\,\mathrm{GB}\) for each 100M dataset and \(20\,\mathrm{GB}\) for SIFT1B. ~\S\ref{sec:scale} describes the multi-CN and multi-MN configurations.

\noindent\textbf{Datasets and systems.}
We evaluate the four public datasets in~\autoref{tab:datasets} against the DM-based d-HNSW~\cite{dhnsw} and DistVS~\cite{distvs}, and the SSD-based DiskANN~\cite{jayaram2019diskann} and PipeANN~\cite{pipeann}. For each dataset, we calibrate \(\alpha\) on 1,000 held-out queries and evaluate on the listed disjoint query set. d-HNSW routes queries through a representative index and fetches selected sub-HNSW partitions from passive-DM, whereas DistVS performs part of the query processing at active memory nodes. DiskANN and PipeANN represent synchronous and pipelined SSD-based graph search, respectively. Because public implementations of d-HNSW and DistVS are unavailable, we implement them following their published designs and configurations.

\noindent\textbf{Resource and index configurations.}
All experiments use a total budget of 32 CPU threads. To ensure fairness, RiftANN dynamically assigns up to 6 CN threads to its shared worker pool, DistVS uses 26 CN and 6 MN threads, and the other systems use 32 CN threads. We additionally use Remote-Vamana and graph-matched in-memory Vamana for breakdown and upper-bound comparisons. Vamana indexes use \((R,L)=(96,128)\) on the 100M datasets and \((128,200)\) on SIFT1B, following prior work~\cite{jayaram2019diskann,pipeann}; d-HNSW and DistVS follow their respective configurations.

\noindent\textbf{Metrics.}
Following prior work~\cite{simhadri2022bigann,pipeann}, we report Recall@10 against mean latency and throughput, using Recall@10 \(=0.90\) as the primary operating point. Selected experiments additionally report 0.95 recall and P99 latency. Unless otherwise stated, each data point is averaged over at least three independent runs.

\begin{table}[t]
\centering
\caption{Evaluation datasets. Each 100M dataset is a subset of its corresponding 1B dataset.}
\small
\renewcommand{\arraystretch}{0.82}
\setlength{\aboverulesep}{1pt}
\setlength{\belowrulesep}{1pt}
\begin{tabular}{@{}lccrc@{}}
\toprule
Name & Size & Type & Dim & Queries \\
\midrule
SIFT100M~\cite{jegou2011sift1b}       & 100M & uint8 & 128 & 10,000 \\
DEEP100M~\cite{babenko2016deep1b}     & 100M & float & 96  & 10,000 \\
SPACEV100M~\cite{simhadri2022bigann}  & 100M & int8  & 100 & 29,316 \\
SIFT1B~\cite{jegou2011sift1b}         & 1B   & uint8 & 128 & 10,000 \\
\bottomrule
\end{tabular}

\label{tab:datasets}
\vspace{-5pt}
\end{table}

  \begin{figure}
      \centering
      \includegraphics[width=0.9\linewidth]{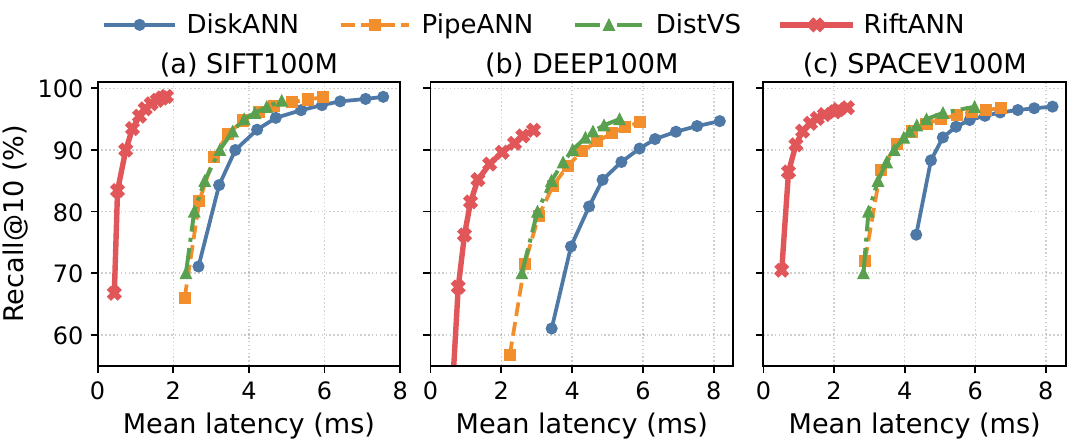}
      \caption{Search latency on 100M-scale datasets.}
      \Description{}
      \label{fig:overall_latency_100m}
      \vspace{-5pt}
  \end{figure}

\vspace{-5pt}
\subsection{Overall Performance}
\label{sec:overall_100m}

\begin{figure}[t]
  \centering
  \includegraphics[width=0.9\linewidth]{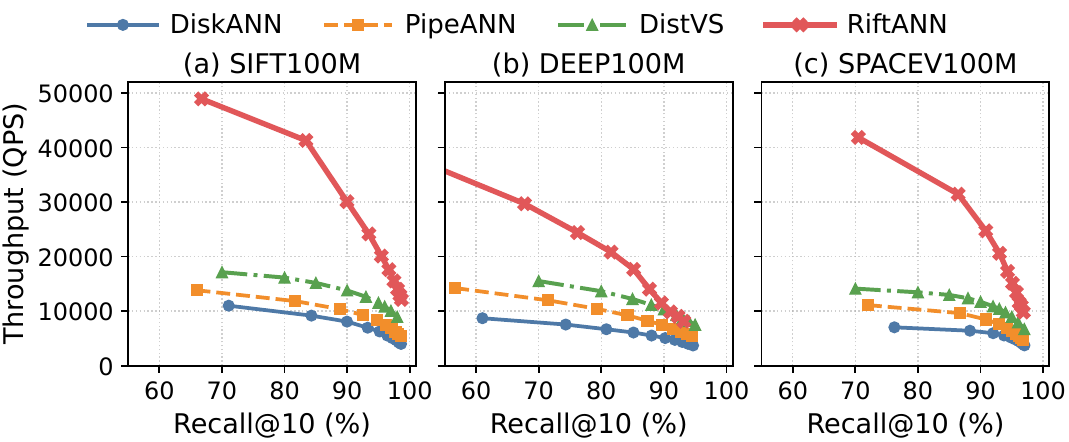}
  \caption{Search throughput on 100M-scale datasets.}
  \Description{}
  \label{fig:overall_throughput_100m}
  \vspace{-5pt}
\end{figure}

Figures~\ref{fig:overall_latency_100m} and~\ref{fig:overall_throughput_100m} compare RiftANN with DiskANN, PipeANN, and DistVS under the same 32-thread CPU budget. We report d-HNSW only in the text because its results fall outside the plotted range.

\noindent\textbf{Search Latency.}
At 0.90 Recall@10, RiftANN achieves mean query latencies of \(0.75\,\mathrm{ms}\), \(2.11\,\mathrm{ms}\), and \(0.88\,\mathrm{ms}\) on SIFT100M, DEEP100M, and SPACEV100M, respectively. These results provide \(1.9\)--\(4.3\times\) speedups over DistVS, \(2.1\)--\(4.3\times\) over PipeANN, and \(2.8\)--\(5.6\times\) over DiskANN. RiftANN remains the lowest-latency system throughout the measured recall range. At 0.90 recall, d-HNSW requires approximately \(620\,\mathrm{ms}\), \(680\,\mathrm{ms}\), and \(570\,\mathrm{ms}\) on the three datasets.

\noindent\textbf{Search Throughput.}
At 0.90 Recall@10, RiftANN sustains \(30.0\,\mathrm{KQPS}\), \(11.0\,\mathrm{KQPS}\), and \(25.9\,\mathrm{KQPS}\), corresponding to \(1.1\)--\(2.2\times\) the throughput of DistVS, \(1.5\)--\(3.0\times\) that of PipeANN, and \(2.2\)--\(4.2\times\) that of DiskANN. d-HNSW remains below \(100\,\mathrm{QPS}\) on all three datasets. RiftANN maintains the highest throughput throughout the measured recall range.

The results reflect how each system maps graph traversal to its data-access architecture. DiskANN and PipeANN remain constrained by block-oriented SSD I/O despite PipeANN's overlap of I/O and computation. d-HNSW amortizes RDMA requests at partition granularity, but cache misses require transferring, deserializing, and searching entire sub-HNSWs. With 32 concurrent query threads, limited partition reuse leaves both latency and network/CPU costs high. DistVS reduces vector transfer through memory-side refinement, but iterative CN--MN coordination places shared-MN computation on the critical path and limits throughput under concurrency. RiftANN instead uses one-sided, record-granular reads and selective exact-vector fetching, avoiding SSD blocks, whole-partition transfers, and repeated memory-side processing. Its asynchronous pipeline overlaps the remaining reads with local evaluation, while impact-aware traversal and shared workers reduce evaluation--expansion stalls and improve compute utilization. These differences explain RiftANN's lower latency and higher throughput under the fixed CPU budget, with its advantage maintained as recall increases. For reference, at 0.90 Recall@10, RiftANN incurs \(1.8\)--\(2.7\times\) the latency of fully in-memory Vamana while sustaining 26--39\% of its throughput.

\vspace{-5pt}
\subsection{Billion-Scale Performance}
\label{sec:billion_scale}

\begin{figure}  
    \centering
    \includegraphics[width=1\linewidth]{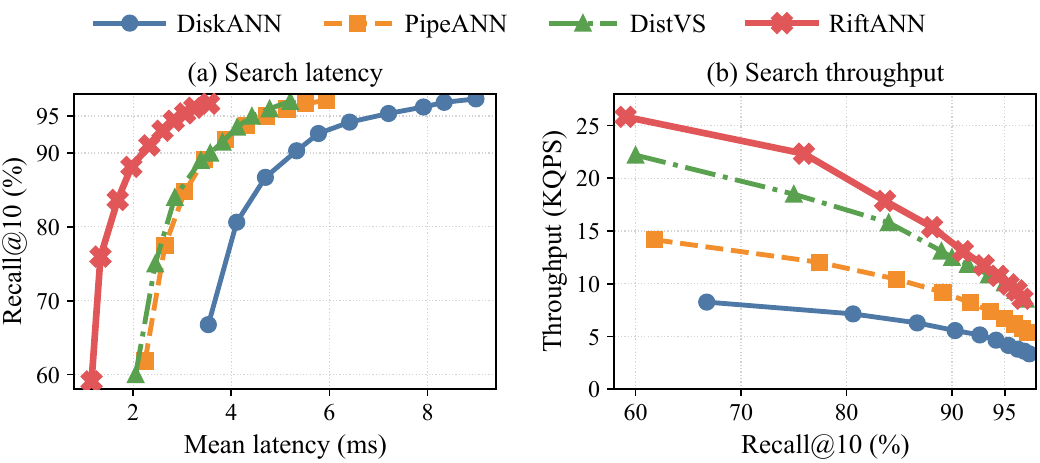}
    \caption{Search performance on the 1B-scale dataset.} 
    \Description{}
    \label{fig:overall_1b}
\end{figure}

We evaluate SIFT1B to determine whether RiftANN's advantage persists as the dataset grows by \(10\times\). We use the same system configurations and aggregate 32-thread CPU budget as in~\S\ref{sec:overall_100m}. d-HNSW is excluded because its SIFT1B index exceeds the \(768\,\mathrm{GB}\) capacity of our MN. \autoref{fig:overall_1b} reports the results.

\noindent\textbf{Search Latency.}
At 0.90 Recall@10, RiftANN achieves \(2.20\,\mathrm{ms}\) latency, compared with \(3.57\,\mathrm{ms}\) for DistVS, \(3.59\,\mathrm{ms}\) for PipeANN, and \(5.28\,\mathrm{ms}\) for DiskANN, providing \(1.62\times\), \(1.63\times\), and \(2.40\times\) speedups, respectively. At 0.95 recall, RiftANN retains \(1.49\)--\(2.34\times\) speedups with \(2.97\,\mathrm{ms}\) latency. Its latency increases by only \(2.93\times\) from SIFT100M, despite the \(10\times\) larger dataset. RiftANN retains its compact navigation state at the CN and accesses remote data at record granularity, so its cost grows with the additional records visited rather than the total index size. In contrast, deeper traversal produces more SSD accesses for DiskANN and PipeANN and more memory-side refinement rounds for DistVS.

\noindent\textbf{Search Throughput.}
At 0.90 Recall@10, RiftANN sustains \(13.9\,\mathrm{KQPS}\), compared with \(12.5\,\mathrm{KQPS}\) for DistVS, \(8.86\,\mathrm{KQPS}\) for PipeANN, and \(5.62\,\mathrm{KQPS}\) for DiskANN. These results correspond to improvements of \(1.11\times\), \(1.57\times\), and \(2.47\times\), respectively. At 0.95 recall, RiftANN sustains \(10.2\,\mathrm{KQPS}\), matching DistVS at \(10.1\,\mathrm{KQPS}\) while outperforming PipeANN and DiskANN by \(1.52\times\) and \(2.38\times\). Overall, RiftANN retains 46\% of its SIFT100M throughput and remains the highest-throughput system across the evaluated recall range.

\subsection{Breakdown Analysis}
\label{sec:breakdown}

\begin{figure}[t]
    \centering
    \includegraphics[width=1\linewidth]{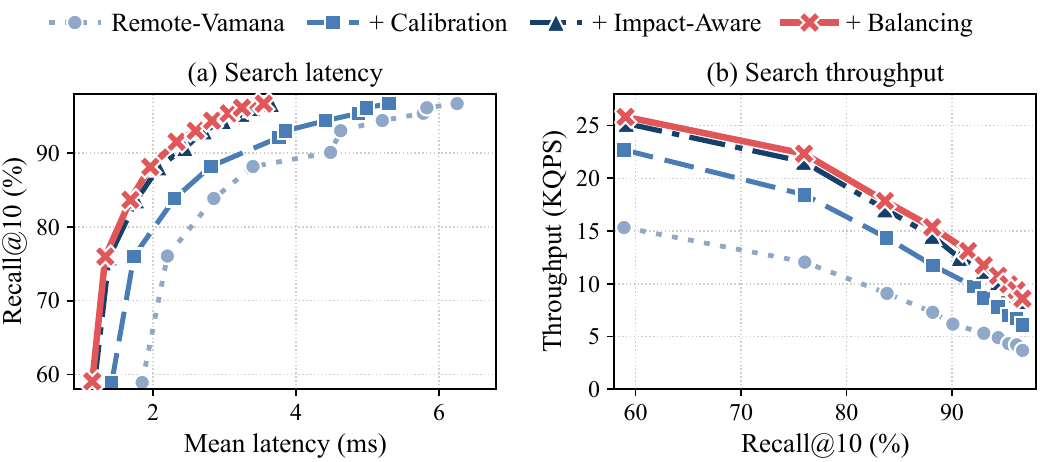}
    \caption{Breakdown analysis of RiftANN.} 
    \label{fig:breakdown}
\end{figure}

We progressively enable RiftANN's three designs over Remote-Vamana on SIFT1B under the same 32-thread CPU budget. \autoref{fig:breakdown} reports the resulting recall--latency and recall--throughput curves. Unless otherwise noted, the following results are measured at 0.90 Recall@10.

\noindent\textbf{Remote-Vamana.}
Remote-Vamana keeps exact-vector fetching and candidate evaluation on the query critical path, resulting in \(4.08\,\mathrm{ms}\) latency and \(6.59\,\mathrm{KQPS}\).

\noindent\textbf{+ Calibrated Pipeline.}
Fetching only required fields, avoiding unnecessary exact-vector reads, and asynchronously issuing the remaining requests reduce latency to \(3.05\,\mathrm{ms}\) and increase throughput to \(10.49\,\mathrm{KQPS}\), improvements of 25.2\% and 59.2\%. The larger throughput gain reflects reduced RNIC contention, although candidate evaluation remains serialized with subsequent expansion.

\noindent\textbf{+ Impact-Aware Traversal.}
Parallel neighbor evaluation and impact-aware expansion further reduce latency to \(2.32\,\mathrm{ms}\) and increase throughput to \(13.18\,\mathrm{KQPS}\), improving them by 23.9\% and 25.6\%. The throughput gain over the calibrated pipeline grows to 33.6\% at 0.95 recall because deeper searches expose more evaluation work that would otherwise remain serialized.

\noindent\textbf{+ Pipeline Balancing.}
Coordinating worker allocation and RDMA concurrency further improves latency and throughput to \(2.20\,\mathrm{ms}\) and \(13.90\,\mathrm{KQPS}\), gains of 5.2\% and 5.5\%. Similar gains at 0.95 recall confirm its consistent benefit. Overall, the three designs reduce latency by 46.1\% and improve throughput by \(2.11\times\) over Remote-Vamana.

\vspace{-5pt}
\subsection{Analysis of Calibrated Pipeline}
\label{sec:eval_calibrated_pipeline}

We evaluate the pipeline's three remote-access decisions, what data to fetch, which candidates to verify, and when to issue the resulting reads, through in-place ablations on SIFT100M. We vary one mechanism at a time while keeping all other settings fixed. Verification uses a fixed search configuration where Verify-all reaches 95\% Recall@10. We calibrate \(\alpha\) using the held-out queries described in \S\ref{sec:setup}.

\noindent\textbf{State-Driven Remote Operations.}
We compare State-driven operations with Full-record, which fetches both the exact vector and adjacency list whenever either is required. Both retrieve 813 exact vectors, expand 405 candidates, and achieve the same recall. By avoiding adjacency lists for verified but unexpanded candidates, State-driven reduces traffic from \(410\) to \(255\,\mathrm{KiB}\) per query, a 37.7\% reduction. Later expansion of 40 verified candidates requires separate adjacency reads, increasing the request count from 813 to 853, a 4.9\% increase. Despite issuing more requests, State-driven increases throughput from \(18.9\) to \(20.1\,\mathrm{KQPS}\), a 6.3\% improvement. The smaller throughput gain relative to the traffic reduction reflects the remaining request-processing and local-computation costs.

\begin{figure}
    \centering
    \includegraphics[height=3.5cm, width=0.9\linewidth]{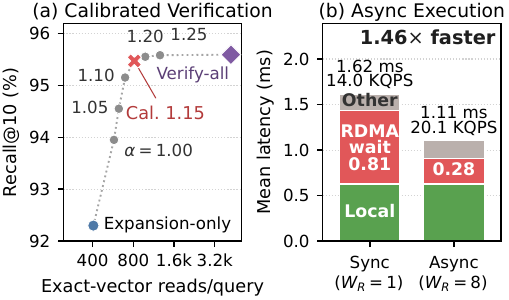}
    \caption{Analysis of the calibrated pipeline.}
    \label{fig:calibrated_pipeline}
\end{figure}

\noindent\textbf{Error-Calibrated Exact Verification.}
\autoref{fig:calibrated_pipeline}(a) compares different verification policies under fixed search settings. Expansion-only reads the 405 expanded candidates and reaches 92.30\% Recall@10. RiftANN selects \(\alpha=1.15\) on separate calibration queries for 99.90\% target coverage. On held-out queries, it achieves 99.88\% coverage, performs 813 exact-vector reads, and reaches 95.47\% recall. Verify-all reads all 4,220 discovered candidates and reaches 95.59\% recall. Calibrated verification therefore eliminates 80.7\% of exact-vector reads with only 0.12 percentage points of recall loss. The selected \(\alpha\) also lies near the trade-off knee. Increasing it from 1.15 to 1.25 adds 447 reads but improves recall by only 0.11 percentage points. DEEP100M and SPACEV100M show similar results: the selected \(\alpha\) values of 1.13 and 1.17 achieve held-out coverages of 99.85\% and 99.83\% against the 99.90\% target, reduce exact-vector reads by 80.9\% and 78.1\%, and incur only 0.13 and 0.16 percentage points of recall loss, respectively. These results show that calibration remains effective across different data distributions.

\noindent\textbf{Asynchronous One-Sided Execution.}
Finally, \autoref{fig:calibrated_pipeline}(b) compares one outstanding read per query with a request window of \(W_R=8\). The two variants perform the same search and remote-access work and achieve nearly identical recall. Asynchronous issuing reduces exposed RDMA wait time from \(0.81\) to \(0.28\,\mathrm{ms}\), a 65.4\% reduction. Consequently, mean latency decreases from \(1.62\) to \(1.11\,\mathrm{ms}\), a \(1.46\times\) speedup, while throughput increases from \(14.0\) to \(20.1\,\mathrm{KQPS}\), a 43.6\% improvement. Because the amount of work is unchanged, these gains result from overlapping RDMA latency with local processing.

\subsection{Analysis of Impact-Aware Traversal}
\label{sec:impact_aware_eval}

We evaluate whether our design overlaps evaluation and expansion without making traversal decisions from an excessively incomplete candidate set. All variants use the same SIFT100M graph, search configuration, and 32-thread CPU budget, differing only in their expansion-admission rules. \textit{WaitAll} requires \(U_q=0\), \textit{U-only} requires \(U_q\le B\), \textit{\(\rho\)-only} requires \(\rho_q\le\theta\), \textit{Impact-aware} requires \(\rho_qU_q\le\epsilon\) and \(U_q\le U_{\max}\), and \textit{Ungated} imposes no condition. All variants merge completed evaluations before each decision and drain outstanding work before termination. We use \(H=64\), \(\epsilon=1\), and \(U_{\max}=128\) by default.

\begin{figure}[t]
  \centering
  \includegraphics[width=\linewidth]{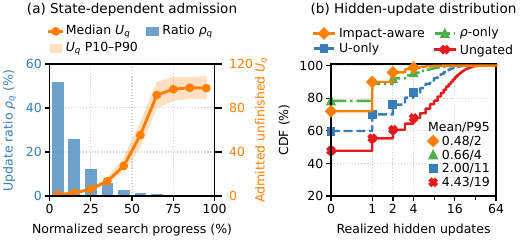}
  \caption{Analysis of Impact-Aware Traversal on SIFT100M. (a) State-dependent expansion admission. (b) Distribution of realized hidden updates across admission policies.}
  \label{fig:impact_aware}
\end{figure}

\noindent\textbf{Expansion Admission.}
\autoref{fig:impact_aware}(a) shows how the gate adapts as the search progresses. During the first 10\% of a search, approximately 52\% of completed evaluations update the candidate set, and the gate admits expansion with a median unfinished backlog of only 1.25. During the final 30\%, the update ratio falls below 0.5\%, allowing nearly 100 evaluations to remain unfinished. The gate is therefore conservative when unfinished work is likely to affect traversal and exposes more overlap as its expected impact decreases.

\autoref{fig:impact_aware}(b) measures the realized impact by counting unfinished evaluations that eventually update the candidate set after each admitted expansion. Impact-aware control limits this number to a mean of 0.48 and a P95 of 2. Across the other admission policies, the mean ranges from 0.66 to 4.43 and the P95 from 4 to 19. Although \(\epsilon\) bounds estimated rather than realized impact, combining update probability with backlog size substantially reduces both the average and tail of the candidate information hidden from traversal.

\noindent\textbf{Admission-Policy Comparison.}
The default Impact-aware configuration achieves 95.47\% Recall@10 at \(1.11\,\mathrm{ms}\) and \(20.1\,\mathrm{KQPS}\), reducing latency by 28.5\% and improving throughput by 28.7\% relative to WaitAll (95.56\%, \(1.55\,\mathrm{ms}\), \(15.6\,\mathrm{KQPS}\)) with only 0.09 percentage points of recall loss. At comparable latency, U-only (\(B=32\)) reaches 94.68\% at \(1.17\,\mathrm{ms}\), and \(\rho\)-only (\(\theta=0.05\)) reaches 95.12\% at \(1.14\,\mathrm{ms}\). Across all swept settings, every U-only and \(\rho\)-only configuration is matched by an Impact-aware configuration with equal or higher recall and equal or lower latency, except \(\rho\)-only with \(\theta=0.2\), which is \(9\,\mu\mathrm{s}\) faster but 1.39 percentage points less accurate. Impact-aware control dominates because \(U_q\) alone does not capture the likelihood that unfinished work changes the candidate set, while \(\rho_q\) alone does not capture how much work remains. Ungated execution reaches \(0.95\,\mathrm{ms}\) but only 92.79\% recall, confirming that uncontrolled overlap is unsafe. Sweeping \(\epsilon\) from 0.25 to 4 reduces latency from \(1.43\) to \(0.99\,\mathrm{ms}\) while recall falls from 95.55\% to 94.96\%; \(\epsilon=1\) lies near the knee.

\noindent\textbf{Robustness across Scale.}
At 0.95 Recall@10, without retuning, the default configuration improves SIFT1B latency from \(4.58\) to \(3.24\,\mathrm{ms}\) (29.3\%) and throughput from \(6.98\) to \(9.49\,\mathrm{KQPS}\) (36.0\%), losing 0.11 percentage points of recall.

\begin{figure}
    \centering
    \includegraphics[height=6.5cm, width=1\linewidth]{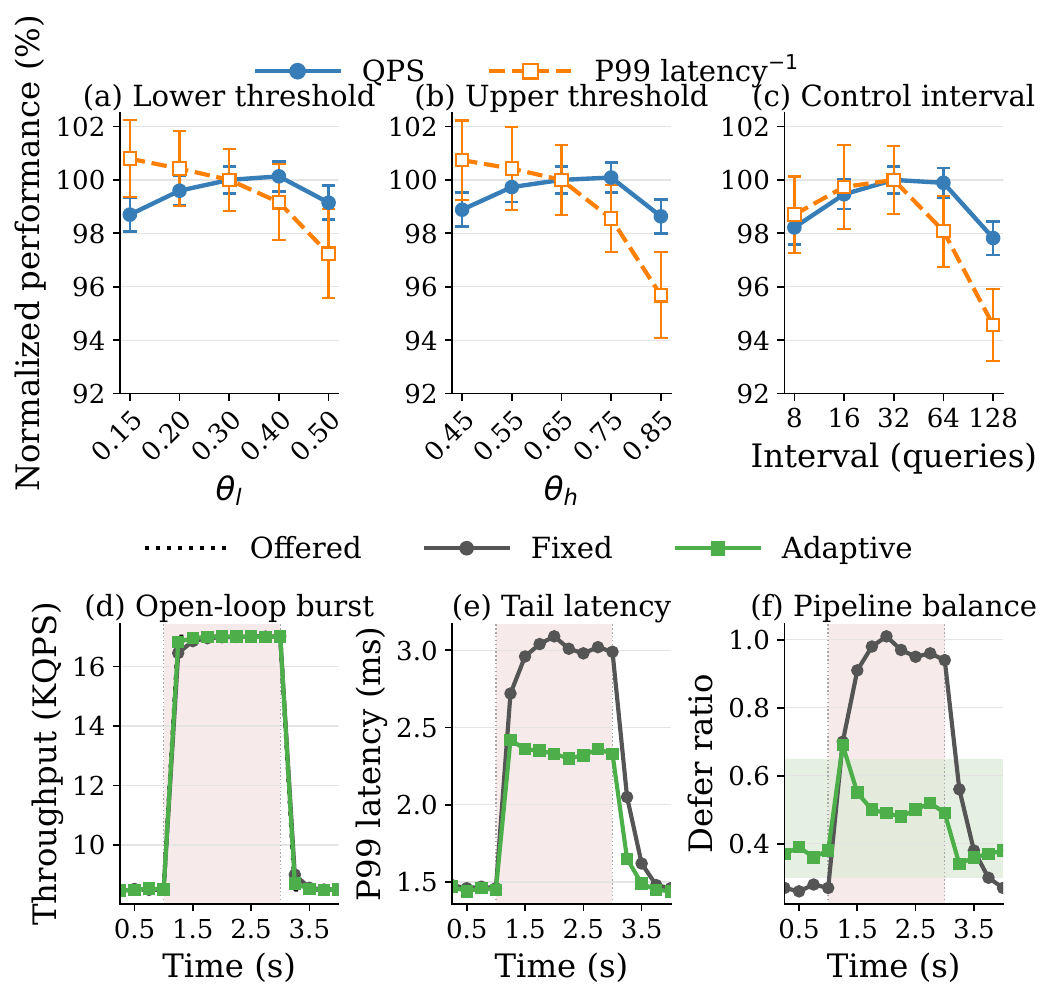}
    \caption{Analysis of Pipeline Balancing.}
    \Description{}
    \label{fig:feedback_pipeline}
\end{figure}

\subsection{Analysis of Pipeline Balancing}
\label{sec:feedback_analysis}

We evaluate the controller's robustness to its parameters and its ability to restore pipeline balance under changing load. All experiments use SIFT100M with the same search configuration and 32-thread CPU budget. We compare the complete controller (\textit{Adaptive}) with \textit{Fixed}, which holds the worker and RDMA request window at their initial values.

\noindent\textbf{Parameter Robustness.}
\autoref{fig:feedback_pipeline}(a)--(c) independently vary \(\theta_{\mathrm{low}}\), \(\theta_{\mathrm{high}}\), and the control interval around their defaults of 0.30, 0.65, and 32 completed expansions. The panels report normalized QPS and inverse P99 latency, for which higher is better, with 95\% confidence intervals over five runs. Across all settings, QPS changes by at most 2.2\%, P99 latency by at most 5.8\%, and Recall@10 by less than 0.02 percentage points. The modest variations follow the expected control behavior. Short intervals add overhead, long intervals react more slowly, and \(\theta_{\mathrm{high}}\) determines the tolerated evaluation backlog, while \(\theta_{\mathrm{low}}\) mainly controls when excess workers are released. The broad plateau around the default shows that the controller does not require narrowly tuned parameters.

\noindent\textbf{Response to Query Bursts.}
As shown in \autoref{fig:feedback_pipeline}(d)--(f), we temporarily double the open-loop query arrival rate while keeping it below system saturation, then return it to the steady level. Reported latency includes request queueing. Before the burst, Adaptive and Fixed exhibit nearly identical behavior. During the burst, Adaptive's defer ratio rises to 0.69 and triggers the controller to increase the mean number of busy evaluation workers from 4.0 to 5.9 while reducing the mean RDMA window from 7.8 to 5.0. These adjustments return the defer ratio to 0.55, whereas Fixed remains between 0.91 and 1.01, indicating persistent evaluation pressure. Adaptive reverses the adjustments after the burst.

Both configurations sustain approximately \(16.9\,\mathrm{KQPS}\). Adaptive reduces P99 latency from \(3.02\) to \(2.35\,\mathrm{ms}\), a 22.2\% reduction, while Recall@10 differs by only 0.01 percentage points.
The controller therefore reduces backlog-induced tail latency by coordinating evaluation capacity and RDMA concurrency, without increasing the CPU budget, dropping requests, or reducing completed throughput.

\vspace{-5pt}
\subsection{Scalability}
\label{sec:scale}

We evaluate SIFT1B on up to four servers connected by a 100-Gbps RDMA switch. A configuration with \(x\) CNs and \(y\) passive MNs is denoted as \texttt{\(x\)C\(y\)M}; each CN uses a 32-thread budget unless otherwise specified. ~\autoref{tab:cluster_scalability} reports single-CN
scale-up and cluster scale-out at matched recall.

At 0.90 recall, increasing the single-CN budget from 16 to 32 threads improves throughput by 40.4\%, whereas 48 threads provide only another 5.0\% and raise latency, placing 32 threads near the scale-up knee. Sharding the index across three passive MNs reduces throughput by only 8.8\% and increases latency by 9.5\%, showing modest memory-side scale-out overhead. Adding CNs is more beneficial: 2C1M and 3C1M achieve \(1.71\times\) and \(2.04\times\) the throughput of 1C1M. Finally, 2C2M improves throughput by 9.2\% and reduces latency by 8.5\% over 2C1M, demonstrating further gains from coordinated CN and MN scaling.

\begin{table}[t]
    \centering
    \caption{SIFT1B scalability at 0.90 Recall@10.
    T/CN denotes the thread budget per CN.}
    \label{tab:cluster_scalability}
    \small
    \renewcommand{\arraystretch}{0.88}
    \setlength{\tabcolsep}{1.5pt}
    \setlength{\arrayrulewidth}{0.3pt}
    \begin{tabular*}{\columnwidth}
    {@{\extracolsep{\fill}}lrrr|rr|rr|r@{}}
        \toprule
        Config.
        & \texttt{1C1M}
        & \texttt{1C1M}
        & \texttt{1C1M}
        & \texttt{1C2M}
        & \texttt{1C3M}
        & \texttt{2C1M}
        & \texttt{3C1M}
        & \texttt{2C2M} \\
        T/CN
        & 16 & 32 & 48 & 32 & 32 & 32 & 32 & 32 \\
        \midrule
        KQPS
        & 9.9 & 13.90 & 14.6 & 13.12 & 12.68
        & 23.75 & 28.36 & 25.94 \\
        Lat. (ms)
        & 1.49 & 2.20 & 3.25 & 2.33 & 2.41
        & 2.58 & 3.24 & 2.36 \\
        \bottomrule
    \end{tabular*}
\end{table}

\vspace{-5pt}
\section{Related Work}
\label{sec:related_work}

\myline{Graph-based and storage-scaled ANNS}
Graph indexes such as HNSW, NSG, and Vamana support high-recall in-memory ANNS~\cite{hnsw,nsg,jayaram2019diskann}, while VSAG further optimizes their execution~\cite{vsag}. Beyond local DRAM capacity, DiskANN, FreshDiskANN, SPANN, SPFresh, Starling, PipeANN, and OdinANN place graph or hybrid indexes on SSDs or tiered storage~\cite{jayaram2019diskann,freshdiskann,spann,spfresh,starling,pipeann,odinann}, optimizing data layout, page-level I/O amortization, pipelining, and update maintenance.

\myline{Distributed and disaggregated vector search}
Distributed systems such as Pyramid and Milvus scale serving across shards~\cite{pyramid,milvus}, while HAKES targets concurrent read-write workloads in a disaggregated architecture~\cite{hakes}. CoTra reduces coordination in RDMA-based distributed graph traversal by tolerating delayed candidate information~\cite{zhi2025towards}. More closely related, d-HNSW combines index caching, RDMA-friendly layout, and query-aware batched loading~\cite{dhnsw}; DistVS combines progressive pruning over multi-tier storage with asynchronous execution, RDMA batching, and decoupled re-ranking~\cite{distvs}; and CXL-ANNS uses graph-aware caching, prefetching, and relaxed dependency execution over CXL-based memory~\cite{jang2023cxlanns}. 

\myline{Relaxed-order execution and latency hiding}
PipeANN relaxes inter-step dependencies to pipeline SSD I/O and computation~\cite{pipeann}, while DistVS decouples re-ranking in a disaggregated setting~\cite{distvs}. More broadly, AIFM and Leap demonstrate application-aware latency hiding and prefetching for far memory~\cite{aifm,leap}. RiftANN builds on these ideas but jointly addresses their limitations.

\section{Conclusion}
\label{sec:conclusion}

We present RiftANN, a graph-based ANNS system for passive-DM that combines calibrated one-sided access, impact-aware asynchronous traversal, and feedback-driven pipeline balancing. RiftANN substantially outperforms the evaluated DM- and SSD-based baselines while narrowing the performance gap to in-memory search.

\bibliographystyle{ACM-Reference-Format}
\bibliography{used-references}

\end{document}